# Transforming Heart Chamber Imaging: Self-Supervised Learning for Whole Heart Reconstruction and Segmentation

Abdul Qayyum*[,1,2], Hao Xu[3], Brian P. Halliday[1,4], Cristobal Rodero[1], Christopher W. Lanyon[5], Richard D. Wilkinson[5], Steven Alexander Niederer[1,2]
[1]National Heart and Lung Institute, Faculty of Medicine, Imperial College London, London, United Kingdom
[2]Turing Research and Innovation Cluster in Digital Twins (TRIC-DT), The Alan Turning Institute, London, UK
[3]College of Mathematical Medicine, Zhejiang Normal University, Zhejiang, Chi-na
[4]Royal Brompton and Harefield Hospital, Guy's  St Thomas' NHS Founda-tion Trust, London, UK
[5]School of Mathematical Sciences University of Nottingham Nottingham, NG7 2RD, UK
Email: a.qayyum@imperial.ac.uk

***Abstract*—Automated segmentation of Cardiac Magnetic Resonance (CMR) plays a pivotal role in efficiently assessing cardiac function, offering rapid clinical evaluations that benefit both healthcare practitioners and patients. While recent research has primarily focused on delineating structures in the short-axis orientation, less attention has been given to long-axis representations, mainly due to the complex nature of structures in this orientation. Performing pixel-wise segmentation of the left ventricular (LV) myocardium and the four cardiac chambers in 2-D steady-state free precession (SSFP) cine sequences is a crucial preprocessing stage for various analyses. However, the challenge lies in the significant variability in contrast, appearance, orientation, and positioning of the heart across different patients, clinical views, scanners, and imaging protocols. Consequently, achieving fully automatic semantic segmentation in this context is notoriously challenging. In recent years, several deep learning models have been proposed to accurately quantify and diagnose cardiac pathologies. These automated tools heavily rely on the accurate segmentation of cardiac structures in magnetic resonance images (MRI) images. Hence, there is a need for new methods to handle such structure's geometrical and textural complexities. We proposed a 2D and 3D two-stage self-supervised deep learning segmentation hybrid transformer and CNN-based architectures for 4CH whole heart segmentation. Accurate segmentation of the ventricles and atria in 4CH views is crucial for analyzing heart health and reconstructing four-chamber meshes, which are essential for estimating various parameters to assess overall heart condition. Our proposed method outperformed state-of-the-art techniques, demonstrating superior performance in this domain.**

## I. Introduction

The global incidence of Cardio Vascular Diseases (CVD) has been steadily rising, reaching an estimated prevalence of 523 million cases worldwide in 2019. This figure represents a nearly twofold increase over a span of 30 years [1]. In comparison to other CVD, ischemic heart diseases stand out as the leading cause of mortality, accounting for more than 9 million deaths in 2021 [2]. In this context, Cardiac Magnetic Resonance Imaging (CMRI) has emerged as a significant aid for the visualization and diagnosis of myocardial diseases. Its proficiency lies in accurately imaging anatomical regions, all while posing minimal risks to the patient [3]. In clinical practice, Multiplanar Cine-MRI sequences play a crucial role in Cardiac Magnetic Resonance Imaging (CMRI), offering the capacity to assess cardiac function and motion across the entire heart. This assessment involves quantification using both short-axis (SAX) and multiple long-axis (LAX) views. Typically, the SAX view acquisition incorporates 8 to 12 spatial slices spanning various cardiac phases. In contrast, each LAX view acquisition captures two, three, and four cardiac chambers within a single spatial slice. The intersections of their respective spatial planes are centered on the left ventricle cavity. Long-axis (LAX) views play a pivotal role in imaging the atria, particularly when compared to short-axis (SAX) views. They are essential for visualizing pathological conditions that affect this specific area of the heart, including malformations and interatrial communications. Moreover, LAX views contribute significantly to the diagnosis of diseases that impact the apical region of the heart [4]. While segmentation for Cardiac Magnetic Resonance Imaging (CMRI) has been extensively investigated, most studies have primarily focused on Short-Axis (SAX) segmentation, with a specific emphasis on utilizing Cine-MRI sequences [5]. Conversely, Long-Axis (LAX) views have received comparatively less attention, and methods developed for other image orientations, particularly SAX [6]–[9], have dominated the research landscape. Limited studies have conducted a comprehensive examination of Cine-MRI long-axis cardiac images, such as the Ω-net method proposed by Vigneault et al. [9].

In this work, the authors addressed multiclass segmentation for SAX, LAX two-chamber, and four-chamber views by employing a predefinition UNet. The bottom features from this UNet were utilized to train a Spatial Transformer Network [10], enabling the learning and estimation of rigid affine transformation matrix parameters to achieve a canonical orientation consistent with clinical practices across the dataset. Following this transformation, the output was directed to a second segmentation pipeline comprising multiple chained UNets. For LAX four-chamber images, the authors annotated over five classes in addition to the background class, enabling whole-heart segmentation. This study involved full cardiac cycle annotation, and the reported results include average Intersection over Union (IoU) scores of 0.856 and 0.845 for LAX two-chamber and LAX four-chamber, respectively.

Another study involving fully annotated LAX four-chamber was conducted by Bai et al. [10]. In this work, the authors compared a semisupervised learning pipeline to a baseline UNet by using a small amount of annotated data. They reached average respective Dice scores of $0.934 \pm 0.029$ and $0.930 \pm 0.032$ by using 200 training slices corresponding to ED and ES timeframes for 100 examinations. Their approach introduces a distinct starting postulate, considering the various axes of view as entangled. The idea is that these axes can be learned through shared anatomical features at their intersections, thereby enhancing segmentation by incorporating surrounding features. The most promising outcomes were attained by employing a multitask pipeline that simultaneously addressed both anatomical intersection detection and segmentation. In previous work, Bai et al. [11] used a VGG16 [12] architecture to automatically segment atria from LAX four- and two-chamber Cine-MRI images. They attained respective Dice scores of $0.93 \pm 0.05$ on the LAX two-chamber left atrium and $0.95 \pm 0.02$ and $0.96 \pm 0.02$ for the LAX four-chamber left atrium and right atrium, respectively. It is worth mentioning that they achieved inference times of approximately 0.2 s for the ED and ES LAX images and 1.4 s when the entire sequence was included.

Several notable works explore the integration of Statistical Models of Deformation (SMOD) with UNet-based segmentation, as proposed by Acero et al. [13]. Their study specifically addressed hypertrophic cardiomyopathy and normal examinations, incorporating Long-Axis (LAX) two- and four-chamber views, with the right atrium being the only missing class. Al Khalil et al. [14] utilized SPADE-GAN image generation along with Variational Autoencoder (VAE)-based label deformation through interpolation over LAX four-chamber images for data augmentation before segmentation. The augmentation generated by the GAN improved their results compared to more traditional morphological alterations. Additionally, Pei et al. [ [14] and Wang et al. [15] leveraged images from the LAX four-chamber view to implement domain adaptation with feature differentiation between Computed Tomography (CT) and Magnetic Resonance (MR) modalities. The M&M2 challenge incorporated LAX four-chamber view images into the segmentation task, with a specific focus on right ventricle segmentation [16]. Participants developed various methods, employing the four-chamber view in independent pipelines [6]–[8] or through shared information pipelines, with LAX data often used to refine Short-Axis (SAX) segmentation. Most other studies have concentrated on Long-Axis (LAX) segmentation, typically targeting one or two anatomical structures. These investigations involve a range of deep learning approaches [3], [17], [18] or hybrid pipelines that combine deep learning, as demonstrated by Zhang et al. [19]. Furthermore, Sinclair et al. [20] and Leng et al. [21] extended their focus to include the LAX three-chamber view orientation that is less frequently examined in segmentation tasks compared to LAX two-chamber and four-chamber views.

## II. Contribution in this paper

1. We developed automatic deep learning models for segmentation of ventricles (LV, RV) , Left Ventricle Myocardium (LVM) and atria(LA, RA) using LAX(2CH,4CH) and SAX (short-axis) views. We have compared our proposed approach on several LAX and SAX datasets.

2. We developed label completion network that was trained on CT dataset and fine-tuned on MRI dataset for whole heart reconstruction. Our method accurately estimates the structure of the whole heart including left ventricle (LV), LV myocardium (LVM), right ventricle (RV), RV myocardium (RVM), left atrium (LA), right atrium (RA), ascending aorta (AA) and pulmonary artery (PA), by taking both LAX and SAX slices.

Fully three-dimensional (3D) networks [11], [12] demonstrate the capability to manage cardiovascular magnetic resonance (CMR) slices with a broad range of variations in slice numbers and orientations. This includes the incorporation of long-axis (LAX) slices, enabling the reconstruction of more precise basal/apical structures, encompassing both atria. Our focus is on addressing the challenge of reconstructing dense whole heart shapes from sparse CMR segmentations.

## III. Dataset used in this study

For model training, two datasets containing Cine-MRI Long-Axis (LAX) 2-chamber and 4-chamber images were utilized, comprising 668 and 1007 examinations, respectively. The variation in dataset sizes is partially attributed to the unavailability of anatomical ground truths for some examinations. In each Cine-MRI study, the end-diastolic (ED) and end-systolic (ES) phases of the cardiac cycle were visually selected and annotated under the supervision of an expert physicist.

The segmentation task in the 2-chamber view focused on three structures: the left ventricle cavity (LVC), the myocardium (MYO), and the cavity of the left atrium (LA). For the 4-chamber view, three additional classes were introduced: the cavities of the right ventricle (RV) and the right atrium (RA), along with the aortic root (AO). The aortic root class was introduced to address cases where the aorta is visible on the image in a 4-chamber orientation. The absence of a consistently present aortic root class across all samples contributes to the segmentation complexity in the 4-chamber view. In both orientations, specific rules based on anatomy were established. The LVC must be disconnected from the background and enveloped by the MYO class. Additionally, the LVC must be in direct contact with the LA at the level of the atrioventricular valve. If present, the AO class must also be in direct contact with the left ventricle cavity at the level of the aortic valve.

TABLE I: Dataset distribution for training and validation of proposed model and state-of-the-art models.

| MRI Views | No of Subjects | Training Data | Validation Data | Dim |
|---|---|---|---|---|
| SAX | 35 | 875 | 450 | 3D |
| 4CH | 35 | 875 | 450 | 2D |
| 2CH | 35 | 875 | 450 | 2D |

### A. TRED-HF dataset

We included 50 cases from the TRED-HF trial [20]. Briefly, patients with dilated cardiomyopathy recovered after pharmacological treatment were included to test a safe way to withdraw treatment. Each patient underwent cine-MRI 1 to 3 times, depending on their remission to heart failure. Since the first scan was the baseline, it is assumed that at that time the heart presents a healthy anatomical presentation. The MRI images included SAX views covering the ventricles and LAX views including 2CH, 3CH and 4CH.

From this dataset we used everything except for 3CH views, focusing on the first scan of each patient. Total number of 135 cases for three-time steps are available without manual annotations. Overall, 53 subject patients were manually annotated to train the model, while the remaining unlabelled images were used to validate and test the model.

We accurately annotated imaging data encompassing 2CH, 4CH, and SAX views across the entire cardiac cycle for a cohort of 53 patients. 35 subjects are used for training and 18 are used for testing the proposed models. Utilizing CVI42 software, we meticulously delineated cardiac structures in both long-axis (LAX) and short-axis (SAX) views. Our annotations covered the left ventricle (LV), myocardium (Myo), right ventricle (RV), left atrium (LA), and right atrium (RA) for the 4CH view. In the 2CH view, we annotated LV, Myo, and LA, while in the SAX view, we focused on LV, Myo, and RV. This comprehensive manual annotation process ensures accurate and detailed segmentation of cardiac structures across different imaging planes and facilitates the development and validation of segmentation algorithms for various cardiac imaging applications.

Table I displays the distribution of subjects with image samples categorized by their views: 4CH (four-chamber), 2CH (two-chamber), and SAX (short-axis). These datasets were utilized for manual segmentation and training our proposed deep learning models. The segmentation and training process are critical stages in developing effective deep learning models for medical image analysis tasks. The distribution of subjects across different views in Table 1 indicates a balanced approach to dataset collection, ensuring representation from various perspectives of cardiac anatomy. This balance is essential for creating models that can accurately segment cardiac structures across different imaging planes.

### B. CMRXmotion challenge dataset

The challenge organizers have made available a dataset comprising multi-contrast, multi-view, multi-slice, and multi-coil Cardiovascular Magnetic Resonance (CMR) imaging data from 300 subjects. The imaging studies encompass cardiac cine and mapping sequences. The dataset, released for the competition, consists of 120 training data, 60 validation data, and 120 test data, with manual segmentations of the myocardium and chambers provided for all subjects.

The collected images include short-axis (SAX), two-chamber (2CH), three-chamber (3CH), and four-chamber (4CH) long-axis (LAX) views. Typically, 5 to 11 slices were acquired for the SAX view, while a single slice was acquired for each of the other views. The cardiac cycle was segmented into 12 to 25 phases with a temporal resolution of approximately 50 ms based on the heart rate. The typical scan parameters included a spatial resolution of $1.5 \times 1.5$ mm$^2$, slice thickness of 8.0 mm, repetition time (TR) of 3.6 ms, and echo time (TE) of 1.6 ms. The parallel imaging acceleration factor was R=3, and signals were acquired with breath-hold.

Manual segmentations of the myocardium and chambers were conducted by an experienced radiologist with more than 5 years of cardiac imaging experience, utilizing ITK-SNAP (version 3.8.0). The segmentation labels and corresponding images were stored in NIFTI format, preserving the original image coordinates. For the LAX cine images, four cardiac chambers were labeled as follows: a) Left atrium - label 1; b) Right atrium - label 2; c) Left ventricle - label 3; d) Right ventricle - label 4. For the SAX cine images, the following labeling was performed: a) Left ventricle - label 1; b) Left ventricular myocardium - label 2; c) Right ventricle - label 3 [21].

### C. The M&Ms-2 Challenge

The M&Ms-2 challenge [16] aimed to reinvigorate and advance research in RV (Right Ventricle) segmentation, addressing data heterogeneity in multi-disease, multi-view, and multi-center cardiac MRI. A unique aspect of the challenge involved incorporating long-axis images to enhance segmentation in the basal plane of the RV, where confusion with the right atrium commonly occurs. The dataset comprised 360 diverse CMR cases from three different vendors and nine scanners, including both short-axis and long-axis 4-chamber views to capture various right and left ventricle pathologies. All images, sourced from three Spanish hospitals, were made openly accessible to facilitate reproducible future research. The dataset was divided into training, validation, and testing sets. Participants were provided with 160 annotated training cases, encompassing both short and long-axis views, and 40 validation cases without annotations. Notably, two pathologies, Dilated Right Ventricle (DRV) and Tricuspid Regurgitation (TRI), were excluded from the training dataset to assess the models' generalization capability to unseen pathologies.

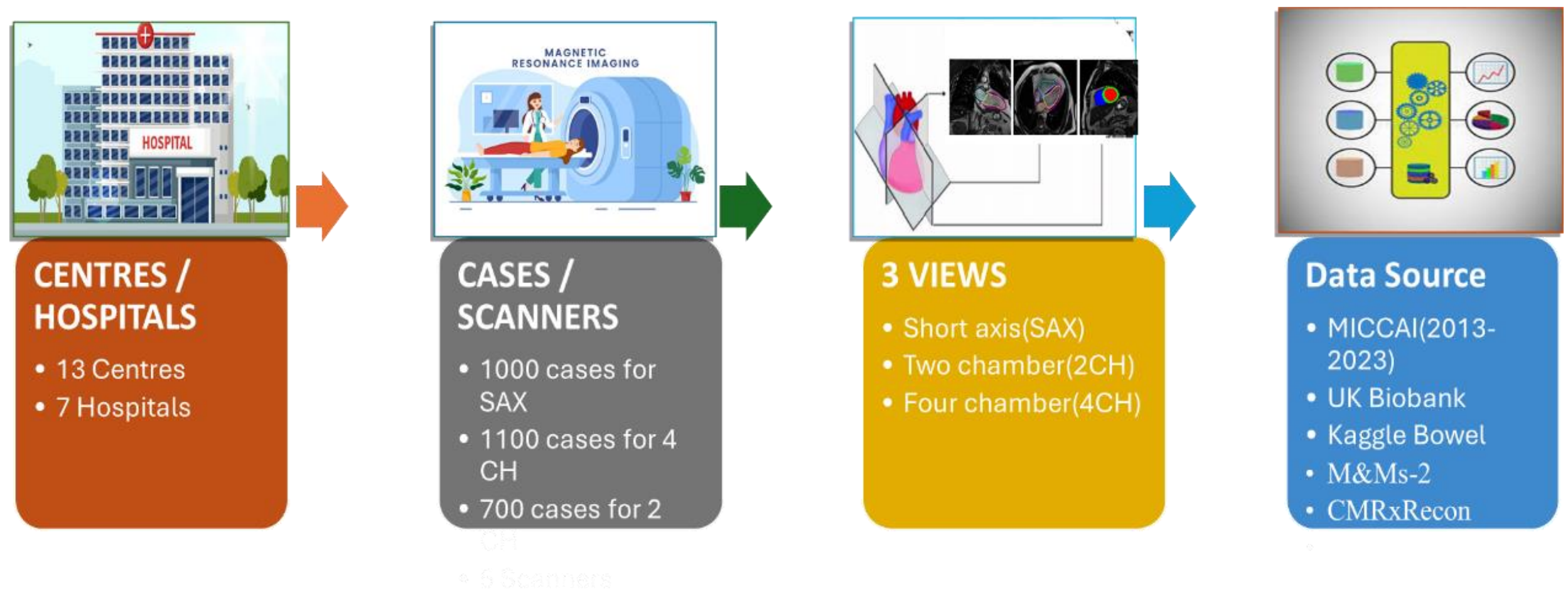

Fig. 1: Various datasets used for SSL model training.

TABLE II: Dataset distribution used for training self-supervised deep learning models

| Dataset | MRI views | No of Subjects | training data | Dim |
|---|---|---|---|---|
| MMs-2 | SAX | 351 | 702 | 3D |
| | 4CH | 351 | 702 | 2D |
| | 2CH | – | – | – |
| CMRxRecon | SAX | 120 | 120 | 3D |
| | 4CH | 103 | 103 | 2D |
| | 2CH | 95 | 95 | 2D |
| Kaggle Bowel | SAX | 116 | 116 | 3D |
| | 4CH | 484 | 484 | 2D |
| | 2CH | 497 | 497 | 2D |
| CAD dataset | SAX | 930 | 930 | 3D |
| | 4CH | 930 | 930 | 2D |
| | 2CH | 930 | 930 | 3D |

### D. *Kaggle Bowel dataset*

The Kaggle Data Science Bowl Cardiac Challenge Data consists of CINE bSSFP cardiac MRI including a short-axis (SAX) stack which was used for ventricular volume quantification. This dataset is publically available [22]. The data was acquired with 8-10 mm slice thickness, spatial resolution between 0.61-1.95 mm x 0.61-1.95 mm, and approximately 30 cardiac frames per slice, at 1.5 and 3 T (MAG-NETOM Aera and Skyra, Siemens Healthcare, Erlangen, Germany). The average distance between consecutive SAX slices was 9.8+/- 0.6. Since the segmentation masks used to generate the EDV and ESV values used as ground truth in the competition were not made publicly available, the entire dataset was re-annotated by an expert observer. The subjects with less than 5 consecutive SAX slices or with the presence of significant motion artefacts were excluded. 491 subject datasets were used for training, 187 for validation and the remaining 424 were reserved for testing. You will examine MRI images from more than 1,000 patients. This data set was compiled by the National Institutes of Health and Children's National Medical Center and is an order of magnitude larger than any cardiac MRI data set released previously.

### E. *Coronary artery disease (CAD) dataset*

In the patient cohort, CAD was confirmed through invasive coronary angiography, considered the diagnostic gold standard. The CMR images were collected using four sequence types: LGE (Late Gadolinium Enhancement), Perfusion, T2-weighted, and SSFP (Steady-State Free Precession). Each sequence includes both long and short axes planes of the heart. The long axis involves two, three, and four-chamber views, with one slice from two different angles captured for each view. The short axis comprises 10 slices from the base to the apex of the heart, with each of the 10 slices captured from 2 different angles.

Based on the provided details, 13 slices have been collected for each patient across the four sequence types. Therefore, each patient contributes a total of $13 \times 4$ CMR images. The calculation of the total number of CMR images (63,648) divided by $13 \times 4$ reveals the participation of 1,224 patients (722 healthy and 502 with CAD) in the creation of this [22].

Table II provides the number of subjects along with the quantity of 2D images for both 2CH (two-chamber) and 4CH (four-chamber) views, as well as the number of 3D samples using SAX (short-axis). These data sets were employed in our self-supervised learning (SSL) approach to train the encoder component of our proposed model. The distribution of 2D images across 2CH and 4CH views, combined with the 3D SAX samples, ensures a diverse and comprehensive training set. This variety is crucial for the SSL framework, as it allows the model to learn robust and generalized features from different perspectives and dimensions. The inclusion of both 2D and 3D data likely enhances the model's ability to understand and represent complex cardiac structures, potentially leading to improved performance in downstream tasks. The structured approach in using these datasets indicates a well-rounded training regimen, leveraging the strengths of both 2D and 3D imaging techniques. The description of datasets used in the SSL part is shown in Figure 1.

## IV. SSL for Medical Image Segmentation

Our proposed model comprised two stages. In the initial step, we trained the model using a large unlabeled dataset through self-supervised learning. Subsequently, in the second stage, we utilized a pretrained encoder and fine-tuned the model with a labeled 2D segmentation model. We elaborated on these two stages in the subsequent sections.

Figure 2 illustrates the mechanism of self-supervised contrastive learning. The crucial step in the contrastive phase involves constructing positive image pairs via data augmentation. We applied various augmentation techniques to randomly selected 2D and 3D foreground patches from 2D and 3D images. Specifically, we utilized two distinct augmented views of the same 2D and 3D patch originating from the same image for contrastive loss. This process functions by bringing the different augmented views closer together if both are generated from the same patch; otherwise, it maximizes disagreement. Contrastive learning provides this functionality on a mini-batch level. We employed masked inpainting, contrastive learning, and rotation prediction as proxy tasks for learning contextual representations of input images. Initially, the network reconstructs the original image, and we applied different classical augmentation techniques such as inpainting, out-painting, and noise augmentation to the image through local pixel shuffling. Following the reconstruction of the original images, we reconstructed both augmented views via regularized contrastive loss with the objective of generating similarity while maximizing agreement. We incorporated a regularized term to adjust the contrastive loss through reconstruction loss dynamically. Finally, multiple patches of dimensions 256×256×32 for 3D and 256×256 for 2D were generated, and various views of a patch were obtained through transformations of the same patches.

## V. Hybrid Parallel Cross Attention (PCAT) Transformer-CNN (MEHTC)

Most recently, there have been many attempts to hybridize the framework consisting of strengths vision transformers and convolutional neural networks. The proposed network consisted of encoder and decoder blocks. In encoder transformer-based module and in decoder the CNN based layers have been proposed. To efficiently segment the ventricles and atrial, in this work, we present an efficient hybrid segmentation framework consisting of the integration of a convolutional neural network and learnable global attention heads. The 3D Hybrid segmentation framework is based on the encoder and decoder block architecture. On the encoder side, we have used swin transformer-based block aided with cross attention window-based mechanism, however, we have used a normal 3DCNN-based module on the decoder side. Following, we have briefly described various components of proposed 3D hybrid segmentation framework which is illustrated in Figure 2.

Unlike the vision transformer (ViT) that computes the relationship between tokens at each step of the self-attentioning module, the Swin transformer is based on the computation of attention within the partition of non-overlapping local windows of lower resolution feature map and original image. In contrast to the original Swin transformer that uses a patch merging layer to empower it for pixel-level tasks, we used rectangular-paralleled-piped windows to accommodate non-square images. To extract different feature maps from each convolutional block in encoder, each block consists of 3D convolutional layers with batch-normalization and ReLU activation functions. We have used the 3D max-pooling layer to reduce the input spatial size of the image. Notice that we have reduced the size of spatial input with the increase in number of layers and we have recovered the spatial resolution in decoder.

Beginning with the top-left voxel, the transformer block employs a regular window partitioning method at each resolution. Feature maps are evenly partitioned into non-overlapping windows. The tokens are combined with features using N-cross head attention. After this, we have computed the cross attention within each transformer window locally. We have used shifted window design to include the connection between neighboring windows (the connection between successive blocks and windowing configuration shifts from the preceding block). We have used. linear layer to project each base window to the query set and query, and each search window to the knowledge set key and value. The cross multi-head attention mechanism computes the attention between both windows and adds it to the base window. Hence the base window gets attention based on corresponding weight information from the searching window which enforces the network to extract feature tokens with corresponding relevance to the input feature set. Following by transformer block, we have applied MLP with GELU on both the output of the base window and searching window before forwarding it to window multicores attention (W-MCA) which enhances the learning ability. In results, the transformer can establish the attention between images and makes it powerful to find the correspondence automatically.

The decoder is based on standard U-net decoder which receives features that have fused both global and local features at each scale from the encoding path along with skip connections.

It also consists of convolution blocks as well as up sampling operation. We have applied 3D up-sampling using bi-linear up-sampling. Finally, the predicted segmentation output (Y $\in$ RH×W) is generated at the final layer of the decoder. In this experiment, we have used 3×3×3 kernel size in both the encoder and decoder and the number of feature map numbers to 16, 32, 64, 128, and 256 for each encoder.

To down-sample the spatial resolution [23]. We have set the kernel size to 2×2×2 for the 3DMaxPool layer on the encoder side. Finally, we have used a transpose3D convolutional layer with stride 2 and 2×2×2 kernel size for up-sampling the size of each decoder. At the end, we concatenated the output of each encoder block to the corresponding decoder block. To generate the final output segmentation map, we have used a 1×1 convolutional layer with softmax function.

We proposed the window-based cross attention mechanism which utilizes multi-size window partitions to limit attention computation in windows. Multi-size window partitions include two different methods, window partition (WP) and window area partition (WAP) [24], to divide the input feature tokens b and s into windows of different sizes. WP partition feature tokens directly into base window set Sba with the size of n×h×w×d and WAP enlarges the window size with the magnifications So the base and searing window size are calculated as:

$$h_{ba}, w_{ba}, d_{ba} = h, w, d \tag{1}$$

$$h_{se}, w_{se}, d_{se} = h, w, d \tag{2}$$

where $h_{ba}, w_{ba}, d_{ba}$ are the size of base windows and $h_{se}$, $w_{se}$, $d_{se}$ are the size of searching windows. To obtain the same amount of two window sets, WAP takes advantage of a sliding window and the stride is set as the base window size, and thus Sse has a size of $n\alpha.\beta.w.\gamma.d$. Through the corresponding windows with different sizes, CAT blocks compute the cross attention between two feature tokens efficiently and avoid large-span searches for precise correspondence. W-MCA to compute the cross attention between acquired base windows and searching windows to find the mutual correspondences. W-MCA is a function mapping a query and a set of key-value pairs to an output, where the query comes from the base windows, and keys and values come from the searching windows. The output is computed as a weighted sum of the values, where the weight assigned to each value is computed by a compatibility function of the query with the corresponding key. W-MCA adopts multi-head attention for ample representation subspaces. W-MCA computes the dot products of query and keys and applies a softmax function to obtain the weights on the values. So, our cross-attention is computed as:

$$W_{mca}(Q_{ba}, K_{se}, V_{se}) = softmax\left(\frac{Q_{ba}K_{se}^{T}}{d}\right)V_{se} \tag{3}$$

$Q_{ba}, K_{se}, V_{se}$ are the query, key, and value metrics.

$Q_{ba} \in R^{nsw}$ is the linear projection of $S_{ba}$ and $K_{se}, V_{se} \in R^{n\mu sw}$) are linear projections of Sse, s=h×w×d $\mu = \alpha.\beta.\gamma$ and c is the dimension of each feature token. The complete process is shown in Figure 3.

### A. Training and optimization of proposed model

We have randomly generated patches with size 128×128×128 from input volume and used different augmentation for data augmentation. The training transforms such as RandCrop, RandGaussianNoise, RandGaussianSmooth, RandShiftIntensity, RandAdjustContrast, and RandZoomd have been used as data augmentation. We have used dice loss and cross-entropy loss functions for training and optimization of the proposed model. The Adam optimizer has been used for training and optimization of the proposed model. The dataset has been normalized using min=0 and max=250.

All volumes have been resampled and interpolated using the same voxel spacing (0.37, 0.53, 0.232).

The sliding window with step size 0.5 has been used to generate the prediction volume with the same patch size (128,128,128). The connected component is used to extract the segmentation's largest components as postprocessing. We have implemented our proposed solution in the PyTorch Framework. The proposed model weights have been trained from scratch and there is no external dataset and pretrained models are used in this proposed solution. We have used all parameters for LAX with same sitting but we used 2D segmentation model.

## VI. 4CH Reconstruction models

### A. Background

The three-dimensional (3D) shape of the heart adjusts in response to sub-clinical disease and risk factors [25], and its alteration has been demonstrated to predict future adverse events. Nevertheless, prevalent imaging techniques like cardiovascular magnetic resonance (CMR) capture only a limited number of slices. The reconstruction of 3D shapes from sparse information could enhance the utility of these studies.

Previous approaches to reconstructing the three-dimensional (3D) shape of the heart from cardiovascular magnetic resonance (CMR) slices have employed various methods, including volume super-resolution [26], [27], point-to-mesh prediction [28], label inpainting [29], and shape deformation [30]. These methods have typically focused on one or two ventricles, utilizing inputs in the form of regular grids or point clouds. Graph convolution networks often output a single mesh per network, and bi-ventricular shape reconstructions may treat ventricular myocardium as a single structure [30] or predict ventricular cavities and myocardium using different networks [28], potentially leading to the neglect of important structures or introducing overlapping between different structures.

Convolutional Neural Networks, on the other hand, can output label maps with different channels, preventing overlap between structures. Anatomically constrained methods [26] usually leverage only short-axis (SAX) CMR slices to simplify the network architecture, either employing a set of 2D networks or one 3D network. However, these methods often require a fixed number of slices and a consistent distance between adjacent slices across the dataset, limiting their applicability to different datasets. Fully three-dimensional (3D) networks [27], [29] exhibit the capability to handle CMR slices with varying numbers and orientations, allowing for the inclusion of long-axis (LAX) slices and enabling more accurate reconstruction of basal/apical structures, including both atria.

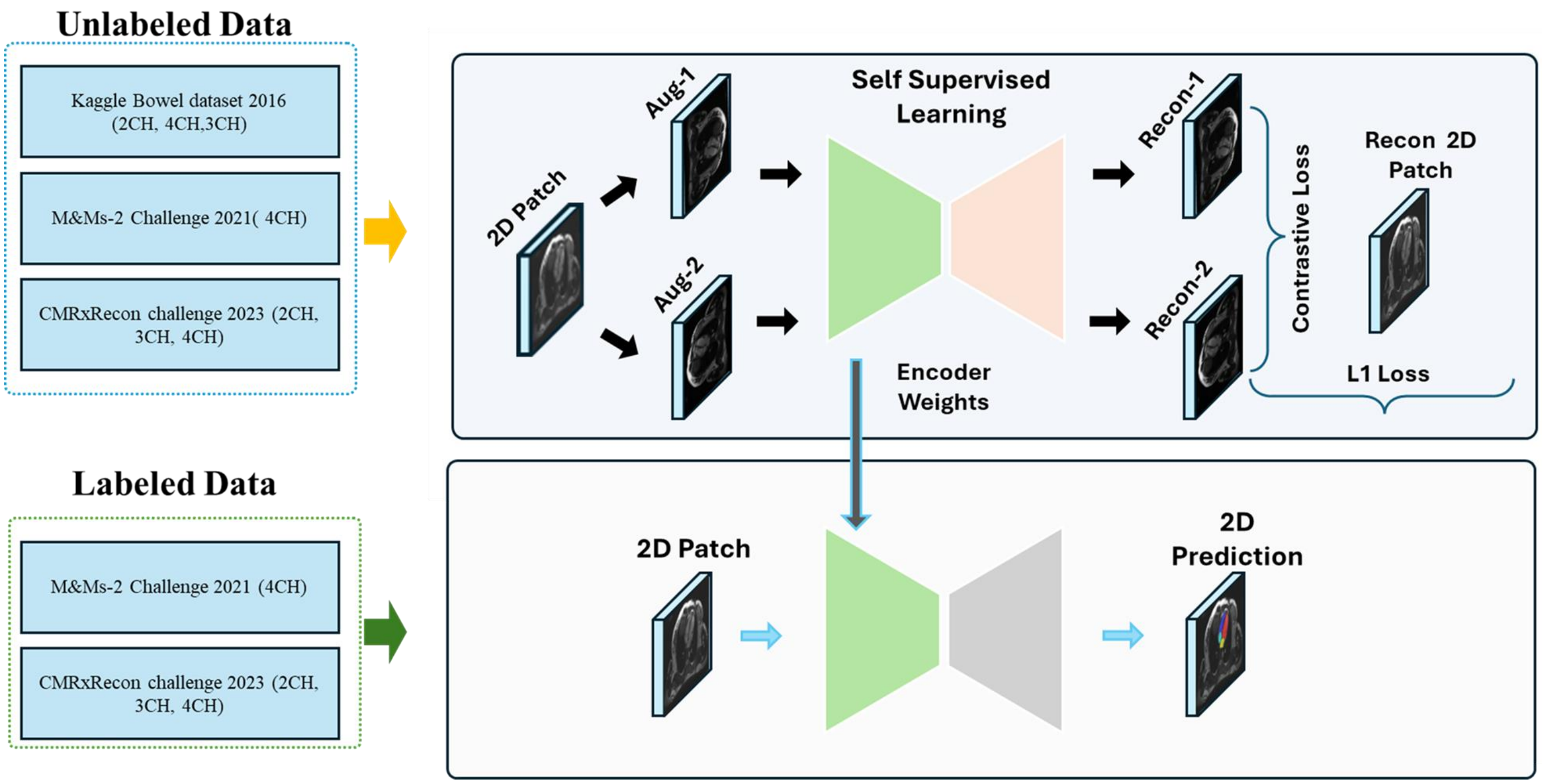

Fig. 2: The proposed self-supervised framework for ventricles and atria segmentation using SAX and LAX datasets

Supervised learning methods typically necessitate paired sparse and dense data for model training. However, such pairs are often unavailable, and sparse data is typically simulated from dense data. Dense data is commonly derived from dense image segmentations and statistical shape atlases. The correction of motion artifacts is essential for improving the plausibility and accuracy of the reconstructed heart shape. This correction is usually integrated into the training stage for supervised learning methods [29] or explicitly applied for unsupervised learning methods.

### *B. 4CH Whole heart Segmentation using label completion network*

The complete pipeline to reconstruct the 4CH whole is shown in Figure 4. The 3D morphology of the atria and ventricles holds significance in studying disease mechanisms. Conventional imaging techniques, such as cardiovascular magnetic resonance (CMR), often capture a limited number of short and long-axis slices. In this study, a label completion U-Net (LC-U-Net) was trained to automatically predict 3D shapes for the ventricles, atria, and valves based on standard CMR views. Dense 3D segmentations from a large coronary computed tomography (CCTA) study were used for training, and the method was tested using simulated short and long-axis CMR slices, considering slice position errors and breath-hold misalignments to mimic actual CMR scans.

The networks were trained and validated using 1700 dense whole heart segmentations obtained from CCTA images, focusing on chamber blood pools and endocardial surfaces while excluding pulmonary veins and the left atrial appendage from the output labels. The cohort comprised patients with suspected coronary artery disease participating in the Scottish Computed Tomography of the HEART (SCOT-HEART) trial. The method accurately estimated the structure of the whole heart, including the left ventricle (LV), LV myocardium (LVM), right ventricle (RV), RV myocardium (RVM), left atrium (LA), right atrium (RA), ascending aorta (AA), and pulmonary artery (PA), by considering both Long-Axis (LAX) and Short-Axis (SAX) slices.

A modified version of the 3D U-Net was employed for label completion, with both input and output being 3D grid volumes of size $160 \times 160 \times 160$ voxels. The input volume represented a sparse label map with seven labels: background (0), LV (1), LVM (2), RV (3), RVM (4), LA (5), and RA (6). The output volume was a dense label map with nine labels: background (0), LV (1), LVM (2), RV (3), RVM (4), LA (5), RA (6), AA (7), and PA (8). Additional details can be found in the referenced source [31].

## VII. Results

We partitioned the dataset into 80% for training and 20% for validation of our proposed model. Our evaluation metrics included Dice coefficient and Hausdorff distance (HD) to assess the performance of both our proposed method and existing state-of-the-art approaches.

In Table III, we present the results obtained from applying our proposed method and existing state-of-the-art techniques on the SAX dataset. We compared our model's performance against well-known architectures such as nnUNet, 3D Res-Net, 3D Dens-Net, and 3D Swin-UNet. For our experiments, we implemented 3D-Res-UNet, Dens-UNet, and Swin-UNet from scratch and trained them on the SAX dataset. We adopted the base setting for nnUNet from its original implementation.

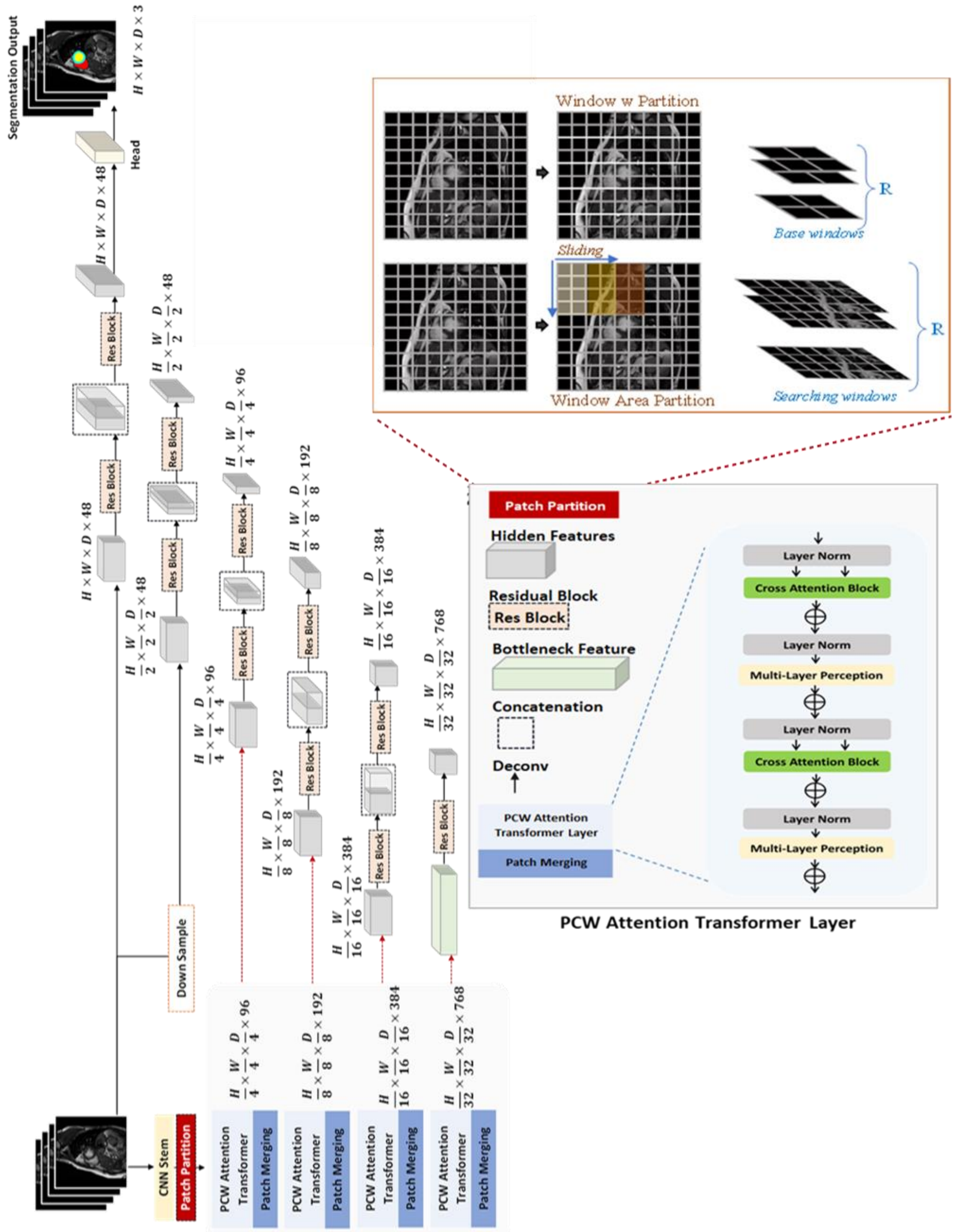

Fig. 3: Proposed model for 2D and 3D segmentation using LAX and SAX datasets.

Our findings revealed notable improvements in performance when employing the self-supervised encoder of our proposed model. Specifically, we observed a significant increase of approximately 4% in the Dice coefficient compared to the nnUNet model. Furthermore, our approach outperformed other established models such as Res-UNet, Dens-UNet, and Swin transformer-based models by achieving a nearly more than 7% enhancement in the Dice coefficient. These results underscore the effectiveness of our proposed method in achieving superior segmentation performance on the SAX dataset, showcasing its potential for advancing state-of-the-art techniques in medical image segmentation.

In Table IV, we observed that our proposed model achieved lower Hausdorff distance (HD) scores compared to existing methods when evaluated on the SAX dataset. A lower HD score indicates that our proposed model generated segmen-tation masks with better accuracy and closer correspondence to the ground truth annotations. The Hausdorff distance is a measure of dissimilarity between two sets of points, in this case, the segmentation masks produced by the model and the ground truth masks. A lower HD score signifies that the distance between corresponding points in the predicted and ground truth masks is smaller, indicating better alignment and agreement between the two. Our model's ability to produce segmentation masks with lower HD scores suggests that it effectively captures the spatial characteristics and fine details of the cardiac anatomy, resulting in more precise delineation of cardiac structures. This improved accuracy is crucial for applications such as medical diagnosis, treatment planning, and image-guided interventions, where precise localization of anatomical landmarks is essential. The superior performance of our proposed model in terms of HD scores highlights its effectiveness in medical image segmentation tasks, particularly in the context of cardiac imaging. By achieving lower HD scores compared to existing methods, our model demonstrates its capability to produce high-quality segmentation results, thereby enhancing the reliability and utility of automated image analysis in clinical practice.

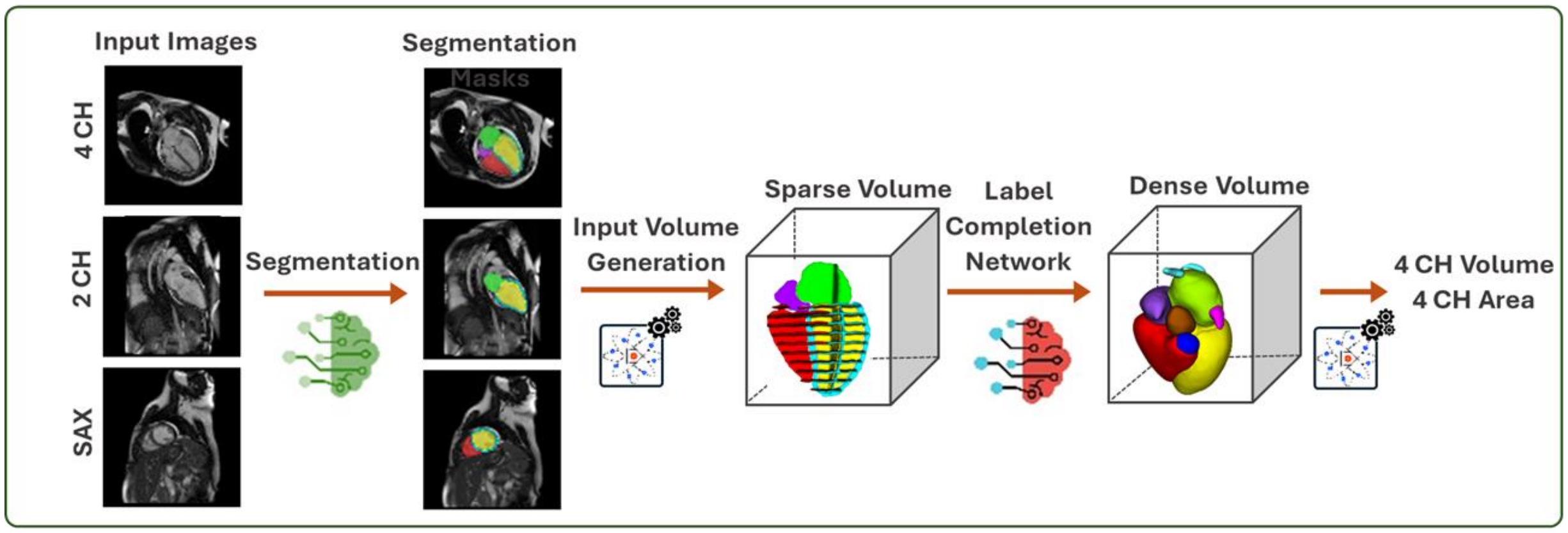

Fig. 4: The 4CH whole heart reconstruction pipeline based on the segmentation of different views of cine MRI images

TABLE III: Dice score for proposed and existing state-of-the art methods using SAX dataset.

| Algorithms | Dice LV | Dice Myo | Dice RV | Avg Dice |
|---|---|---|---|---|
| Proposed SSL | 0.943±0.037 | 0.887±0.031 | 0.906±0.041 | 0.948±0.019 |
| Proposed w/o SSL | 0.913±0.050 | 0.834±0.052 | 0.848±0.059 | 0.915±0.034 |
| nnUNet | 0.909±0.053 | 0.830±0.042 | 0.824±0.063 | 0.904±0.034 |
| DensNet | 0.859±0.102 | 0.781±0.064 | 0.779±0.090 | 0.863±0.066 |
| Res-UNet | 0.824±0.120 | 0.771±0.065 | 0.800±0.091 | 0.856±0.070 |
| Swin-UNet | 0.871±0.075 | 0.746±0.088 | 0.838±0.064 | 0.864±0.059 |

TABLE IV: HD score for proposed and existing state-of-the art methods using SAX dataset

| Algorithms | LV HD | Myo HD | RV HD | Avg HD |
|---|---|---|---|---|
| Proposed SSL | 2.78±1.61 | 2.85±1.75 | 3.72±1.63 | 3.19±1.51 |
| Proposed w/o SSL | 8.07±16.52 | 10.42±21.35 | 8.29±16.80 | 12.21±22.68 |
| nnUNet [32] | 11.36±19.92 | 15.02±22.93 | 11.28±17.84 | 17.05±23.66 |
| Dens-UNet [33] | 10.35±20.64 | 17.28±27.30 | 31.84±32.34 | 31.96±32.31 |
| Res-UNet [34] | 43.46±33.85 | 48.51±34.08 | 32.41±35.87 | 50.76±33.95 |
| Swin-UNet [35] | 46.13±29.61 | 53.38±27.58 | 15.33±23.14 | 54.03±27.01 |

Figures 5 and 6 depict the performance metrics of Dice scores and Hausdorff distances (HD) using the SAX dataset. The performance is presented for all testing subjects in mean and standard deviation boxplots. Our analysis reveals that our proposed method, particularly when with self-supervised learning (SSL), achieves higher Dice scores compared to other methods, indicating superior segmentation accuracy across all tested subjects. Our proposed method with SSL consistently outperforms alternative methods, yielding higher Dice scores for segments involving the left ventricle (LV), myocardium (Myo), and right ventricle (RV), as well as for the overall average Dice score. This demonstrates the effectiveness of our approach in accurately delineating cardiac structures and achieving robust segmentation performance across diverse testing subjects. Figure 6 illustrates the HD scores, which quantify the dissimilarity between corresponding points in predicted and ground truth masks. Lower HD scores signify better spatial alignment and closer agreement between segmentation results and ground truth annotations. Our proposed method exhibits lower HD scores, indicating superior spatial accuracy and better capture of anatomical details compared to alternative methods. Notably, Res-UNet and Swin-transformer models tend to produce higher HD scores, particularly in segments involving the LV, Myo, and RV. Overall, our findings underscore the efficacy of our proposed method, particularly when leveraging SSL, in achieving superior segmentation performance on the SAX dataset across all tested subjects.

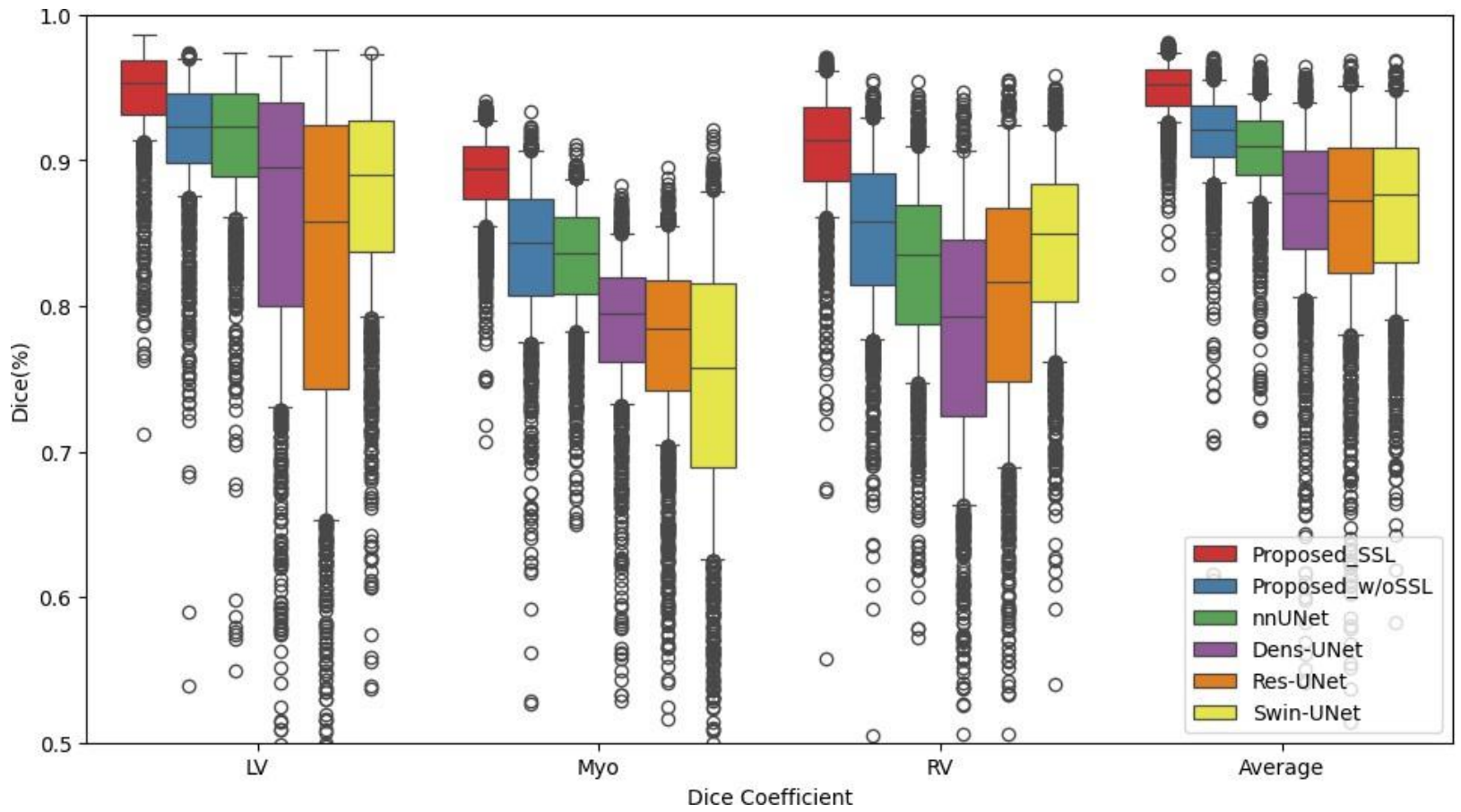

Fig. 5: Dice performance using SAX based on proposed and SOTA models.

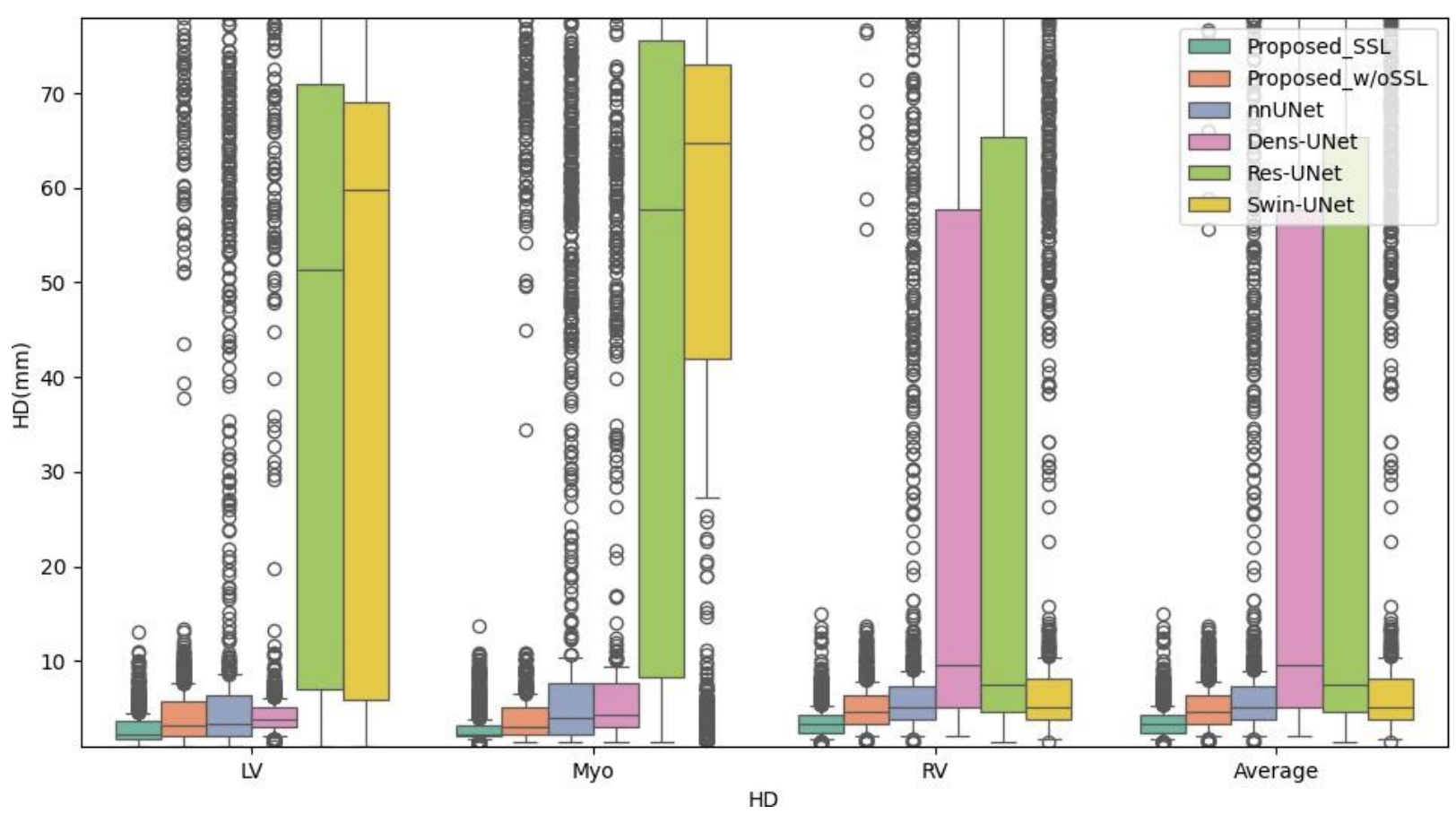

Fig. 6: HD performance using SAX based on proposed and SOTA models.

Our research proposed two different 2D segmentation models for left and right ventricles and atrial segmentation on the 4CH and 2CH datasets. One proposed variant incorporates self-supervised learning, leveraging a large unlabeled dataset, while the other does not employ self-supervised techniques.

Our investigation demonstrates that our self-supervised approach yields higher Dice scores compared to existing state-of-the-art (SOTA) methods. Specifically, our proposed solution achieves superior segmentation performance, as evidenced by its high Dice scores when compared to other deep learning models. The incorporation of self-supervised learning enables our model to leverage additional unlabeled data effectively. This allows the model to learn rich and robust feature representations, leading to improved segmentation accuracy. By exploiting self-supervised learning techniques, our model gains a deeper understanding of the underlying structure and variability present in the data, thereby enhancing its segmentation capabilities. Furthermore, our analysis reveals that our proposed solution outperforms other deep learning models commonly used in medical image segmentation tasks. This superior performance underscores the effectiveness of our approach in accurately delineating cardiac structures from 4CH and 2CH datasets. Overall, our research demonstrates the effectiveness of self-supervised learning in improving segmentation performance on cardiac imaging datasets. By achieving higher Dice scores compared to existing methods and other deep learning models, our proposed solution holds promise for advancing the state-of-the-art in cardiac image segmentation.

Table V and VI present the Dice scores and Hausdorff distances (HD) obtained from our segmentation experiments using the 4CH dataset. Similarly, Table VII and VIII display the corresponding results for the 2CH dataset. In the 4CH dataset, we segmented the left ventricle (LV), myocardium (Myo), right ventricle (RV), left atrium (LA), and right atrium (RA). Conversely, for the CH2 dataset, our segmentation focused on the left ventricle (LV), myocardium (Myo), and left atrium (LA). Our analysis reveals notable performance metrics across both datasets. The Dice scores reflect the overlap between the predicted segmentation masks and the ground truth annotations, with higher scores indicating better agreement. Conversely, the Hausdorff distances quantify the dissimilarity between corresponding points in the predicted and ground truth masks, with lower distances signifying closer alignment.

TABLE V: Dice using proposed and state-of-the-art methods based on 4CH.

| Algorithm | Dice LV | Dice Myo | Dice RV | Dice LV | Dice LA | Avg Dice |
|---|---|---|---|---|---|---|
| Proposed SSL | 0.974±0.009 | 0.906±0.023 | 0.934±0.035 | 0.952±0.023 | 0.936±0.047 | 0.963±0.012 |
| Proposed w/o SSL | 0.975±0.009 | 0.903±0.027 | 0.931±0.033 | 0.949±0.035 | 0.899±0.120 | 0.954±0.029 |
| nnUNet [32] | 0.900±0.044 | 0.742±0.074 | 0.883±0.059 | 0.918±0.061 | 0.922±0.032 | 0.939±0.017 |
| Dens-UNet [33] | 0.897±0.056 | 0.753±0.075 | 0.865±00.070 | 0.878±0.148 | 0.872±0.114 | 0.927±0.053 |
| Res-UNet [34] | 0.883±0.087 | 0.737±0.115 | 0.861±0.079 | 0.828±00.284 | 0.831±0.216 | 0.913±00.078 |
| Swin-UNet [35] | 0.874±0.095 | 0.739±0.110 | 0.862±0.072 | 0.717±.342 | 0.837±0.238 | 0.882±0.112 |

TABLE VI: HD for proposed and state-of-the-art Methods using 4CH.

| Algorithm | HD LV | Hd Myo | HD RV | HD LV | HD LA | Avg HD |
|---|---|---|---|---|---|---|
| Proposed SSL | 2.42±4.07 | 8.07±014.88 | 3.97±3.63 | 6.97±15.44 | 11.17±24.30 | 6.52±16.03 |
| Proposed w/o SSL | 5.56±13.69 | 11.37±15.43 | 6.94±11.67 | 24.45±44.56 | 0.899±0.120 | 10.94±24.17 |
| nnUNet [32] | 7.39±13.36 | 11.66±2.55 | 6.09±4.09 | 8.05±018.48 | 13.53±32.30 | 9.34±18.03 |
| Dens-UNet [33] | 7.34±0.056 | 12.36±3.48 | 9.95±11.36 | 29.18±46.57 | 19.33±37.13 | 15.63±28.75 |
| Res-UNet [34] | 8.76±11.68 | 12.70±4.23 | 11.87±18.68 | 17.91±34.04 | 25.14±47.46; | 15.27±28.58 |
| Swin-UNet [35] | 7.77±7.61 | 13.14±4.97 | 7.91±6.54 | 34.92±48.79 | 28.71±48.80 | 18.49±33.21 |

TABLE VII: Dice using proposed and state-of-the-art methods based on 2CH.

| Algorithm | Dice LV | Dice Myo | Dice LA | Avg Dice |
|---|---|---|---|---|
| Proposed SSL | 0.958±0.060 | 0.896±0.053 | 0.940±0.057 | 0.931±0.040 |
| Proposed w/o SSL | 0.957±0.056 | 0.901±0.031 | 0.912±0.095 | 0.923±0.043 |
| nnUNet [32] | 0.948±0.070 | 0.897±0.040 | 0.923±0.091; | 0.923±0.057 |
| Dens-UNet [33] | 0.924±0.057 | 0.814±0.069 | 0.903±0.082 | 0.880±0.060 |
| Res-UNet [34] | 0.940±0.020 | 0.831±0.051 | 0.922±0.061 | 0.897±0.031 |
| Swin-UNet [35] | 0.941±0.021 | 0.828±0.057 | 0.930±0.031 | 0.900±0.028 |

TABLE VIII: HD using proposed and state-of-the-art methods based on 2CH.

| Algorithm | HD LV | HD Myo | HD LA | Avg HD |
|---|---|---|---|---|
| Proposed SSL | 3.33±1.51 | 10.32±7.71 | 4.38±3.23 | 3.95±5.72 |
| Proposed w/o SSL | 3.43±1.19 | 9.79±4.82 | 9.27±19.53 | 5.29±8.04 |
| nnUNet [32] | 6.47±11.39 | 12.16±10.20 | 13.30±25.26 | 8.94±15.62 |
| Dens-UNet [33] | 10.62±23.64 | 11.69±16.62 | 25.18±34.96 | 18.31±27.87 |
| Res-UNet [34] | 13.12±30.28 | 17.66±27.49 | 16.23±27.43 | 15.34±27.67 |
| Swin-UNet [35] | 8.42±26.82 | 14.09±25.81 | 10.05±20.62 | 11.98±26.92 |

By segmenting multiple cardiac structures in the 4CH dataset and focusing on key structures in the 2CH dataset, our proposed model demonstrates its versatility and effectiveness in capturing anatomical variations across different imaging modalities. Moreover, the achieved Dice scores highlight the accuracy and reliability of our segmentation approach, surpassing existing methods and yielding promising results for clinical applications. Furthermore, the relatively low Hausdorff distances attained in both datasets underscore the precision of our segmentation method, as it effectively captures the spatial details and contours of cardiac structures. This enhanced spatial accuracy is crucial for tasks such as surgical planning, disease diagnosis, and treatment evaluation, where precise delineation of anatomical boundaries is essential. Overall, our segmentation results on the 4CH and 2CH datasets showcase the robustness and effectiveness of our proposed model, demonstrating its potential for advancing cardiac image analysis and improving patient care in clinical settings.

Figures 7 and 8 present the performance comparison between our proposed method and state-of-the-art (SOTA) models using the 4CH dataset. The evaluation is conducted using testing subjects, focusing on segmentation results for the left ventricle (LV), myocardium (Myo), right ventricle (RV), left atrium (LA), and right atrium (RA). Our analysis reveals that both variants of our proposed method, with and without self-supervised learning (SSL), achieve better Dice scores compared to SOTA models, as depicted in Figure 3. Specifically, our proposed solutions consistently exhibit higher Dice scores across all cardiac structures, indicating superior segmentation accuracy. This superiority is particularly pronounced in segments involving the RV, LA, and RA. Notably, the Swin transformer-based model demonstrates suboptimal performance, especially in segments related to the RV, LA, and RA. This observation suggests potential challenges in distinguishing between the RV and RA structures, possibly due to segmentation confusion. Furthermore, Figure 8 illustrates the Hausdorff distances (HD), which quantify the dissimilarity between predicted and ground truth masks. Our proposed methods consistently achieve lower HD scores compared to SOTA models, indicating better spatial alignment and closer agreement with ground truth annotations.

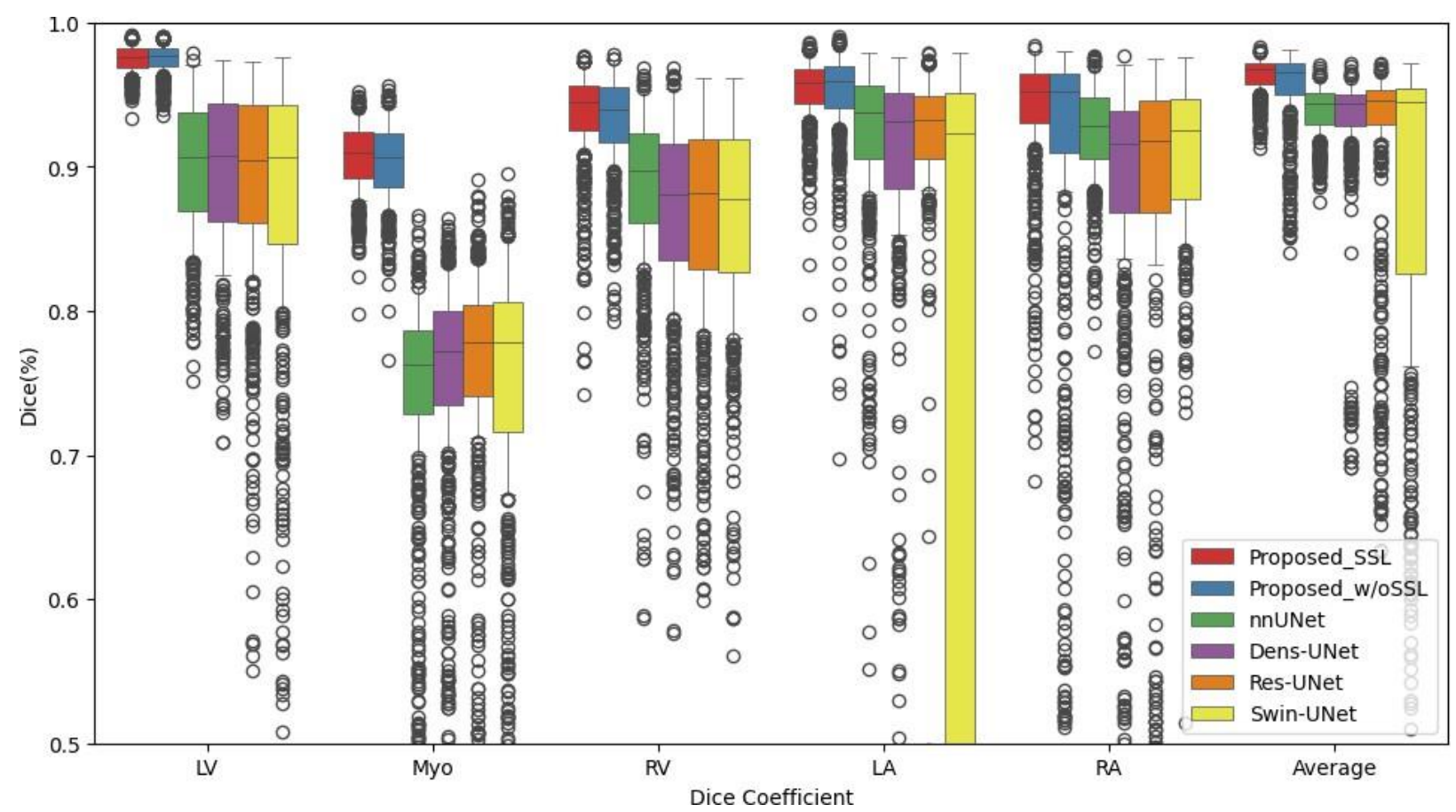

Fig. 7: Dice performance using CH4 based on proposed and SOTA models.

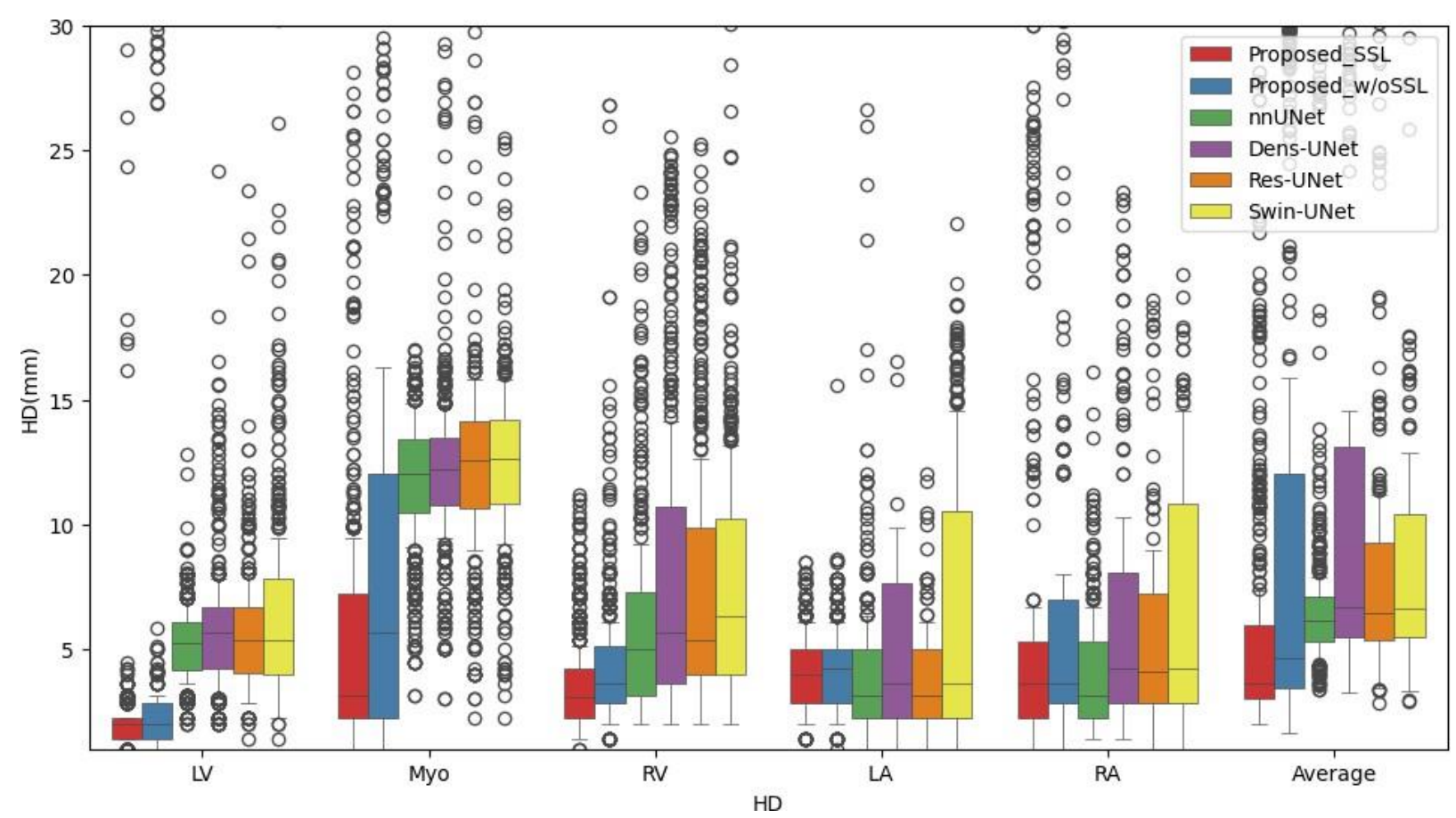

Fig. 8: HD performance using CH4 based on proposed and SOTA models.

In Figures 9 and 10, we present the performance evaluation of our proposed model using the 2CH dataset, focusing on segmentation results for the left ventricle (LV), myocardium (Myo), and left atrium (LA). Our analysis reveals that our proposed model achieves competitive Dice scores compared to state-of-the-art (SOTA) models, as shown in Figure 5. Across all evaluated structures, including LV, Myo, and LA, our proposed model consistently demonstrates robust segmentation accuracy, showcasing its effectiveness in accurately delineating cardiac structures. However, when examining the Hausdorff distances (HD) in Figure 6, we observe that neither our proposed model nor the SOTA model achieves superior performance in segmenting the right ventricle (RV). Specifically, the HD scores for RV segmentation are higher compared to other structures, indicating challenges in accurately capturing the spatial details and contours of the RV. This observation suggests potential difficulties in distinguishing between the RV and other adjacent structures, leading to suboptimal segmentation performance in this region. Overall, while our proposed model achieves competitive segmentation results for LV, Myo, and LA segmentation on the 2CH dataset, there are challenges in accurately segmenting the RV. Future improvements in segmentation algorithms may focus on addressing these challenges to enhance the overall performance of cardiac image segmentation, particularly in regions with complex anatomical structures such as the RV.

Figure 11 showcases the segmentation results of the left ventricle (LV), myocardium (Myo), and right ventricle (RV) using short-axis (SAX) views for three different subjects. The subjects are depicted with base, middle, and apex slices. The proposed model demonstrates superior performance in the base slices, producing segmentation maps that closely resemble the ground truth (GT). Both the proposed model and state-of-the-art (SOTA) models perform well in the middle slices. However, the SOTA models encounter segmentation errors in the apex slices, likely due to the smaller organ areas present in these slices. The proposed model shows the best performance in subjects 2 and 23, while it performs the worst in the more challenging subject 22. In comparison, most SOTA models struggle with both base and apex slices, unlike the proposed self-supervised model, which maintains better performance as illustrated in Figure 11.

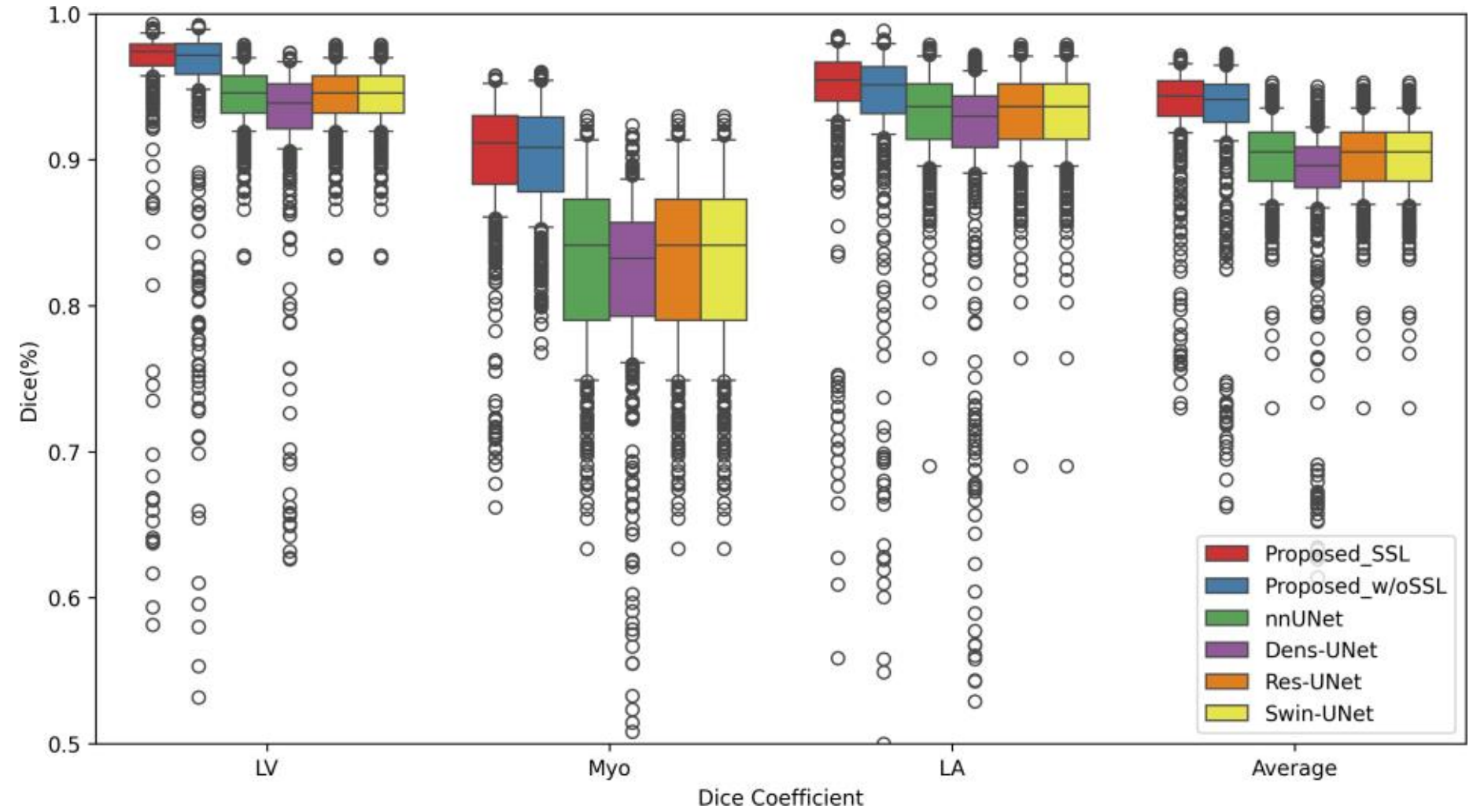

Fig. 9: Dice performance using CH2 based on proposed and SOTA models.

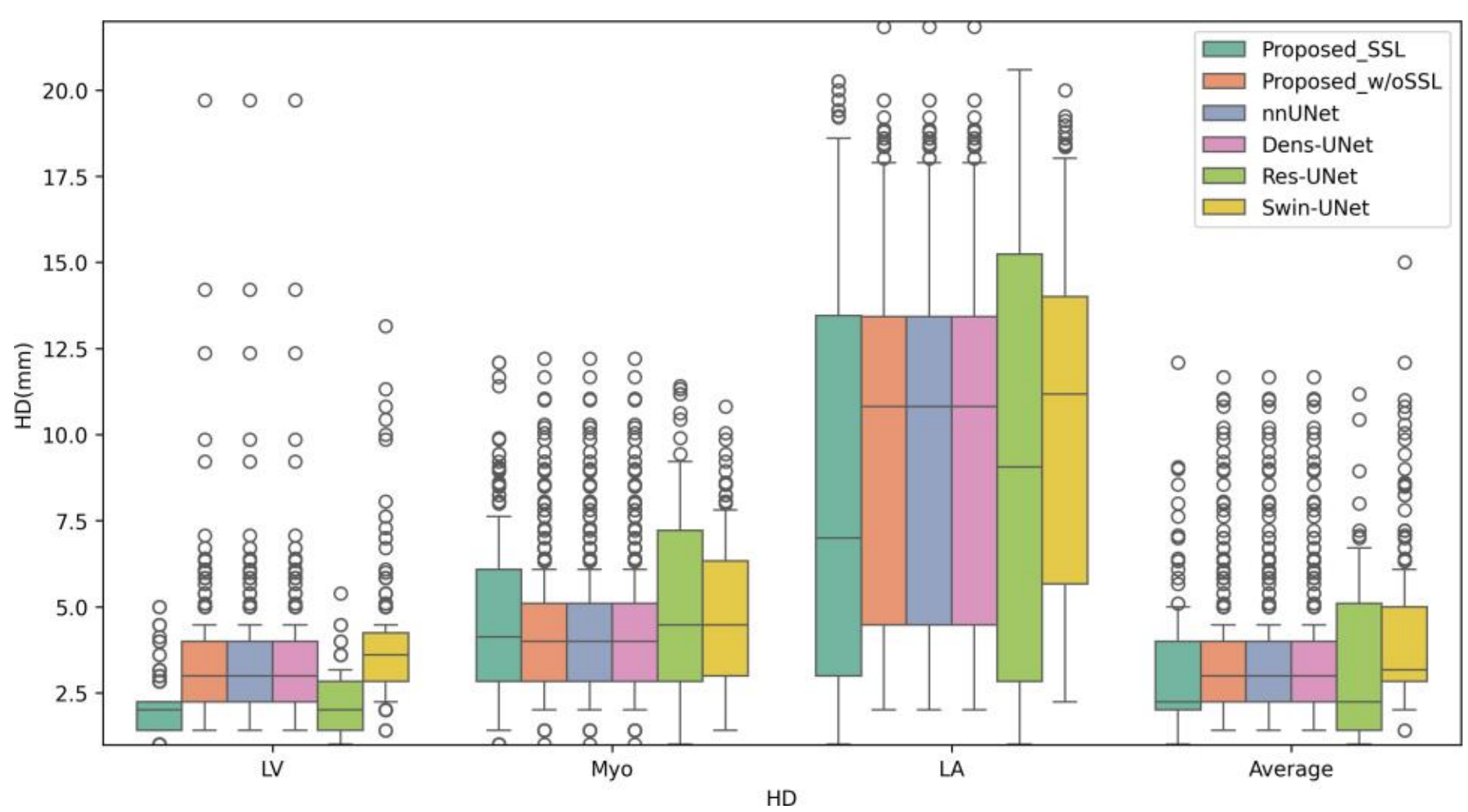

Fig. 10: HD performance using CH2 based on proposed and SOTA models.

The proposed model achieves better segmentation performance for base slices, closely aligning with the ground truth. This suggests robust handling of larger and more distinct cardiac structures in the base region. Both the proposed and SOTA models deliver high-quality segmentation in middle slices. This indicates that these slices, with moderate complexity, are well-handled by current segmentation models.

SOTA models struggle with segmentation in the apex slices, showing errors due to the smaller and less distinct cardiac structures. The proposed model, while also challenged, performs comparatively better in this difficult region. The proposed model excels with subjects 2 and 23, indicating strong generalization capabilities across different anatomical variations. Subject 22 presents more difficulty, resulting in the worst performance from the proposed model, highlighting areas for potential improvement. The proposed self-supervised model consistently outperforms SOTA models in base and apex slices, which are typically more challenging. This superior performance underscores the effectiveness of the self-supervised approach in handling a wider range of slice complexities in SAX views. Figure 3 demonstrates that the proposed self-supervised model provides more accurate and consistent segmentation across various SAX slices and subjects, particularly excelling where SOTA models fall short.

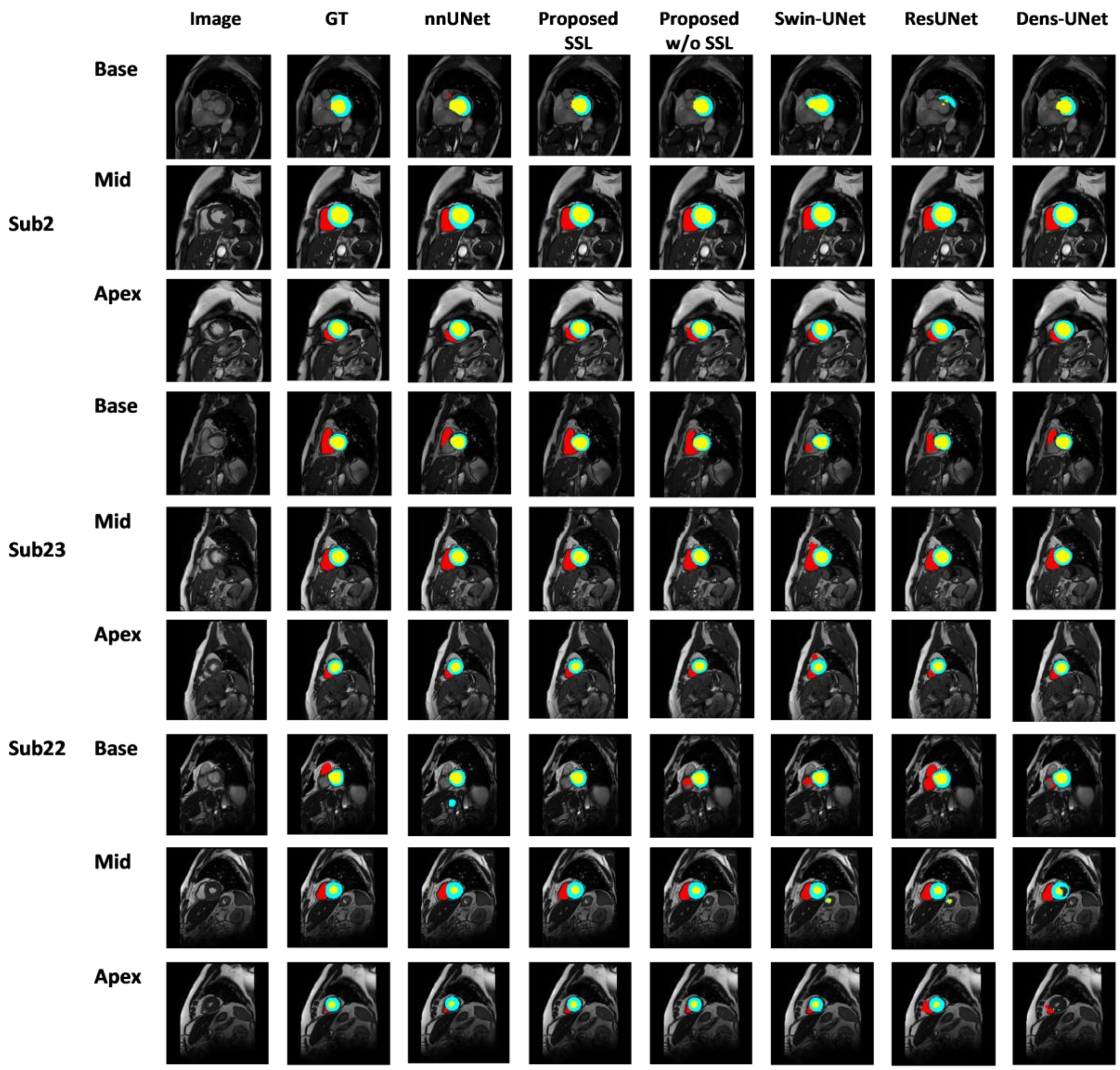

Fig. 11: presents a visual comparison of the proposed model and state-of-the-art (SOTA) models using the SAX dataset for segmenting the left ventricle (LV), myocardium (Myo), and right ventricle (RV). The comparison involves three test subjects, with each model being tested on base, middle, and apex 2D slices for each subject.

Figure 12 shows a comparison of the proposed model and state-of-the-art (SOTA) models using the 2CH dataset for segmenting the left ventricle (LV), myocardium (Myo), and left atrium (LA). Figure 12 illustrates the segmentation of the left ventricle (LV), myocardium (Myo), and left atrium (LA) in a 2CH view using the proposed model and several state-of-the-art (SOTA) models. We tested our model on three different subjects: subjects 2, 22, and 23. The proposed semi-supervised learning (SSL) model achieved superior visual performance compared to the SOTA models, as demonstrated in Figure.

The SSL-based model consistently produces clear and accurate segmentation masks for LV, Myo, and LA across all tested subjects. The visual quality of the segmentations is high, indicating that the model can effectively delineate the cardiac structures. The proposed SSL model outperforms the SOTA models in visual clarity and segmentation accuracy. In subjects 2, 22, and 23, the SSL model demonstrates better boundary definition and less segmentation overlap or gaps compared to SOTA models. The enhanced performance of the SSL model suggests improved learning and generalization capabilities, likely due to the effective utilization of both labeled and unlabeled data. The visual results indicate that the SSL model can handle the variability in cardiac anatomy among different subjects more effectively than the SOTA models. Overall, the proposed SSL model shows significant improvement in visual segmentation performance for the 2CH view, highlighting its potential as a robust tool for cardiac image analysis.

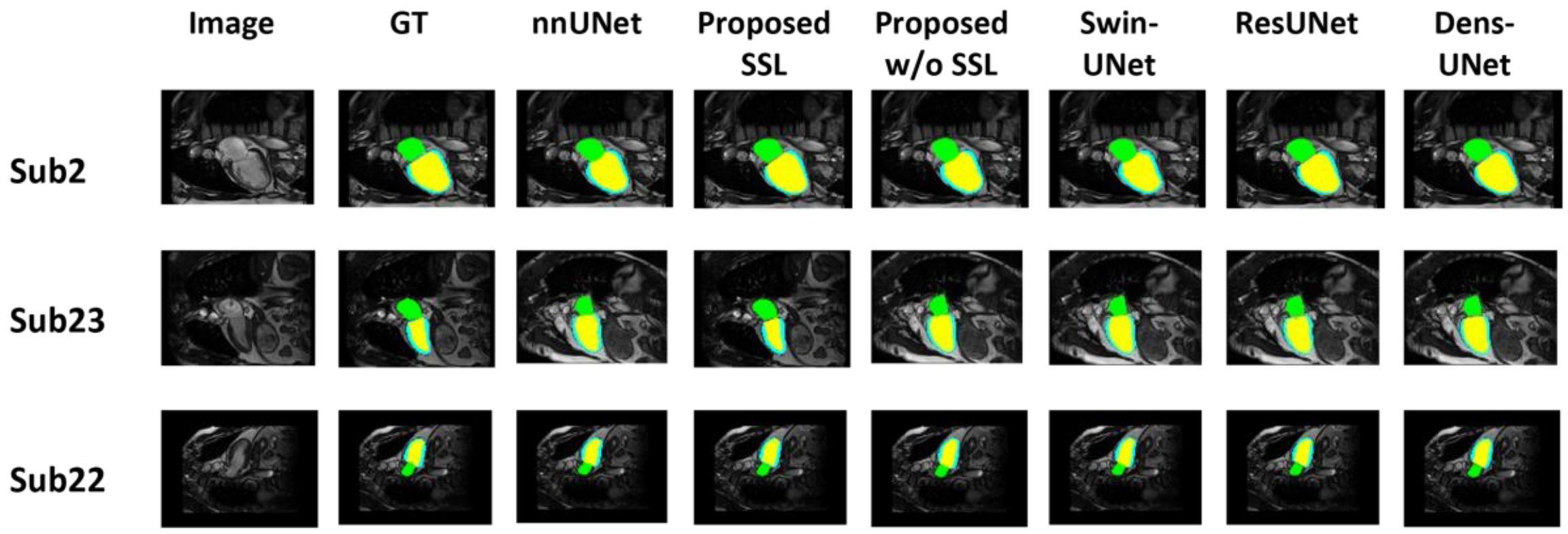

Fig. 12: shows a comparison of the proposed model and state-of-the-art (SOTA) models using the 2CH dataset for segmenting the left ventricle (LV), myocardium (Myo), and left atrium (LA).

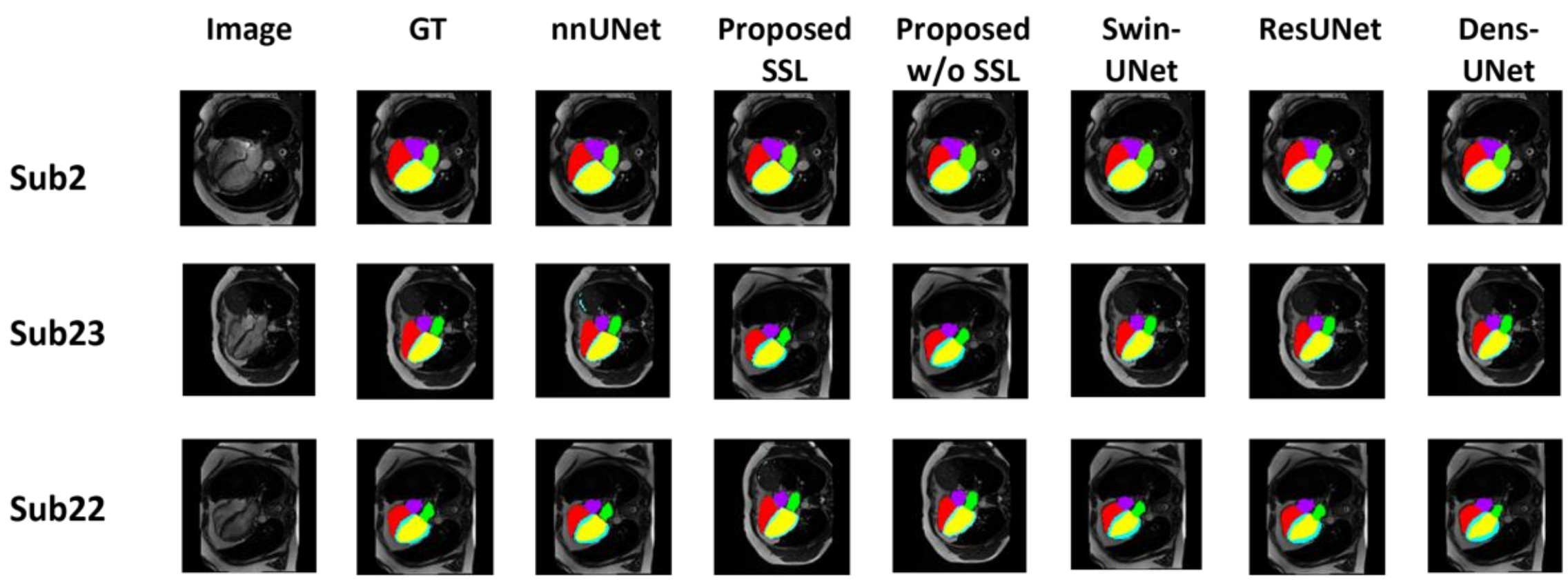

Fig. 13: displays the results based on the 4CH dataset, comparing the proposed model and state-of-the-art (SOTA) models for segmenting the left ventricle (LV), myocardium (Myo), right ventricle (RV), left atrium (LA), and right atrium (RA) using three different test subjects.

Figure 13 displays the results based on the 4CH dataset, comparing the proposed model and state-of-the-art (SOTA) models for segmenting the left ventricle (LV), myocardium (Myo), right ventricle (RV), left atrium (LA), and right atrium (RA) using three different test subjects. Figure 13 presents the visualizations of segmentation masks for LV (left ventricle), Myo (myocardium), RV (right ventricle), LA (left atrium), and RA (right atrium) using three different subjects from the 4CH test dataset. The proposed model generates segmentation masks that are visually like those produced by other state-of-the-art methods, which also exhibit optimal performance. Notably, there is a small gap between the ventricles and atria when using the Dense-UNet and Swin Transformer models, indicating precise segmentation. However, the Res-UNet model tends to overestimate the segmentation, particularly evident in subjects 22 and 23, as depicted in Figure. This overestimation may result in less accurate segmentation, particularly in distinguishing boundaries between adjacent cardiac structures. The proposed model's segmentation closely matches the ground truth, indicating high accuracy in visual segmentation. The Res-UNet model requires further refinement to reduce overestimation and improve segmentation accuracy.

Figure 14 presents visualizations of the whole-heart reconstruction of four chambers using our label completion and segmentation model. The first row showcases results based on ground truth, while the second row displays the outcomes of our proposed model with SSL (Self-Supervised Learning), and the third row exhibits the results without SSL. Subsequent rows illustrate results obtained using state-of-the-art (SOTA) models. Observations reveal instances of false pixel predictions in both the 3D Res-UNet and DensUNet methods. Specifically, in the Res-UNet method, subject 22 exhibits poor performance, with false predictions observed in both the right ventricle (RV) and left ventricle (LV). Similarly, the Dens-UNet method demonstrates low performance in subject 23, as depicted in Figure 14.

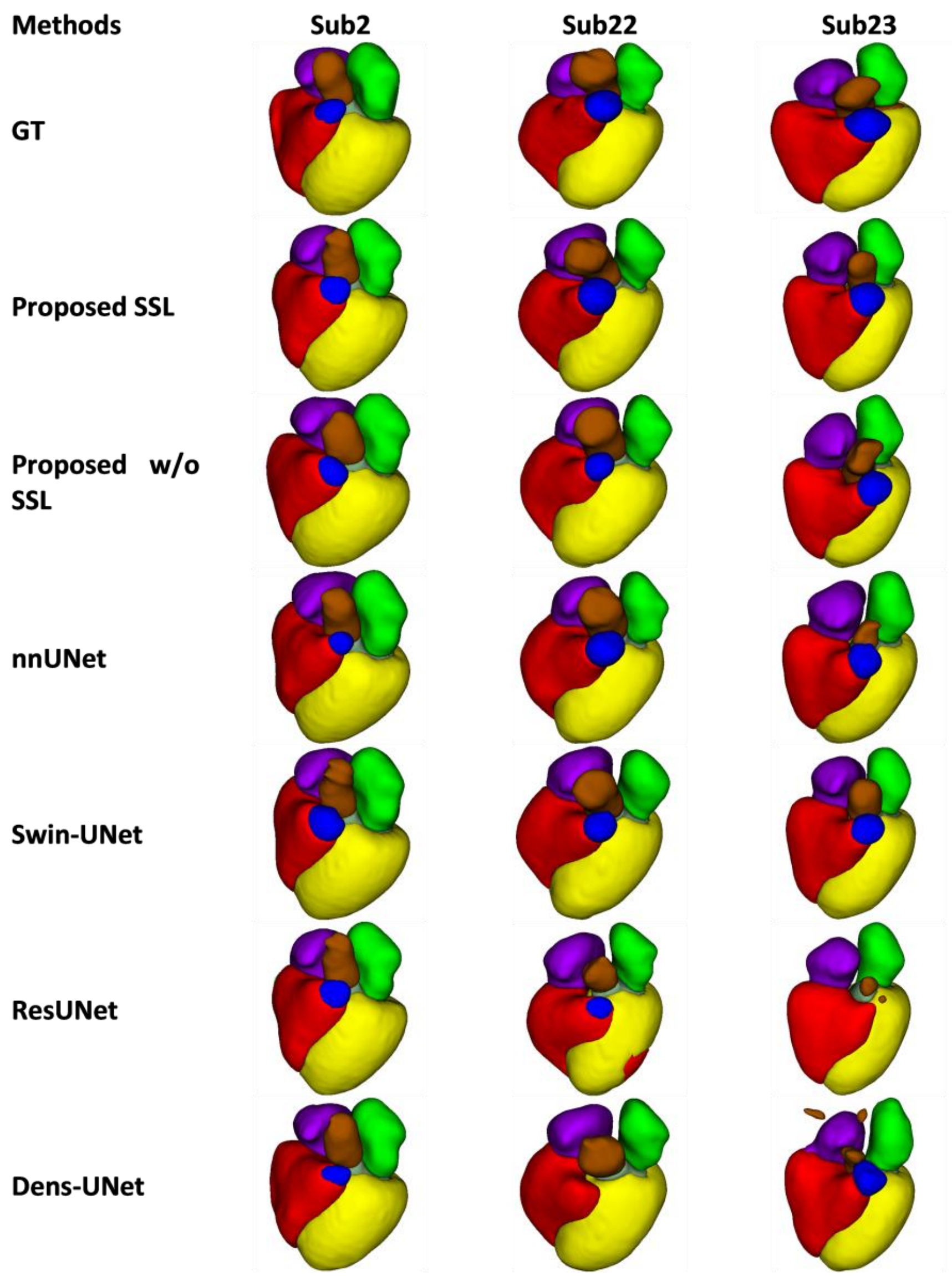

Fig. 14: Displays the visualizations of the reconstructed four-chamber heart using both the proposed model and state-of-the-art (SOTA) models, evaluated on the test subjects sub02, sub21, and sub22.

This analysis underscores the importance of understanding and addressing the limitations and challenges associated with different reconstruction methods, particularly in accurately capturing the intricate details of cardiac structures.

## VIII. CONCLUSION

We proposed a deep learning model for 4CH heart segmentation and reconstruction. The network comprises self-supervised learning (SSL) and downstream modules. Our proposed solution demonstrates potential for accurate segmentation and comprehensive 4CH whole heart reconstructions. The network integrates self-supervised learning and downstream modules, suggesting a robust architecture designed to leverage both pre-training on unlabeled data and fine-tuning on labeled data. Training on a large dataset of over 1000 unlabeled samples indicate a significant effort to ensure the model learns diverse and representative features, enhancing generalizability and performance. The SSL-based decoder's superior performance compared to SOTA models highlights the effectiveness of the self-supervised approach in improving segmentation accuracy. The model's capability for accurate 4CH heart segmentation and reconstruction makes it valuable for clinical applications, such as analyzing heart health and aiding in the diagnosis and treatment of cardiac conditions.

## ACKNOWLEDGMENT

Thanks to our collaborators who provided the data used in this work.